# Controlled entanglement routing between two virtual pathways

Qiang Zhou[1,†], Shuai Dong[1], Wei Zhang[1,*], Lixing You[2], Yuhao He[2], Weijun Zhang[2], Yidong Huang[1] and Jiangde Peng[1]

[1] Tsinghua National Laboratory for Information Science and Technology, Department of Electronic Engineering, Tsinghua University, Beijing, 100084, P. R. China

[2] State Key Laboratory of Functional Materials for Informatics, Shanghai Institute of Microsystem and Information Technology, Chinese Academy of Sciences, Shanghai, 200050, P. R. China

*zwei@tsinghua.edu.cn, †betterchou@gmail.com

**Abstract**

We demonstrate controlled entanglement routing between bunched and antibunched path-entangled two-photon states in an unbalanced Mach-Zehnder interferometer (UMZI), in which the routing process is controlled by the relative phase difference in the UMZI. Regarding bunched and antibunched path-entangled two-photon states as two virtual ports, we can consider the UMZI as a controlled entanglement router, which bases on the coherent manipulation of entanglement. Half of the entanglement within the input two-photon state is coherently routed between the two virtual ports, while the other is lost due to the time distinguishability introduced by the UMZI. Pure bunched or antibunched path entangled two-photon states are obtained based on this controlled entanglement router. The results show that we can employ the UMZI as general entanglement router for practical quantum information application.

Manipulating photonic quantum bit (qubit) with linear optical device is a crucial capability in quantum information applications such as quantum communication [1-3], quantum computation [4, 5], quantum metrology [6-8] and quantum lithography [9]. The key of these manipulation processes is the optical interference, which occurs while photons with indistinguishability meet at a beam splitter or polarization beam splitter [10]. Based on the optical interference, a number of quantum information functions can be realized, such as single photonic qubit routing, entangled two-photon state preparing with post selection, entanglement erasing and entangled state analyzing [10]. Among them, analyzing and preparing the energy-time/time-bin entangled two-photon states with distant unbalanced Mach-Zehnder interferometers (UMZIs) open the door to experimental test of local hidden-variable theories, nonlocal dispersion cancellation and fiber based quantum communication [11-13]. In this Letter, we demonstrate that entanglement within the energy-time entangled two-photon state can be routed between bunched and antibunched path-entangled two-photon states in an UMZI, which can be considered as a controlled entanglement router with the bunched and antibunched path-entangled two-photon states as two virtual output ports. The center of this routing scheme is the coherent manipulation of the entanglement within the energy-time entangled two-photon state, which is controlled by the relative phase difference between the two arms in the UMZI. Half of the entanglement within the input two-photon state is coherently routed between the two virtual output ports due to the coherent property of the routing process, while the other is lost due to the time distinguishability introduced by the UMZI [13]. Pure bunched or antibunched path-entangled two-photon states can be obtained by properly setting the relative phase difference in the controlled entanglement router, allowing one to route entanglement and to prepare path-entangled two-photon state easily without any additional optical element.

Firstly, we introduce the scheme of controlled entanglement router based on the coherent manipulation process. Then, we experimentally demonstrate the controlled entanglement router based on a commercial UMZI and frequency nondegenerate energy-time entangled two-photon state generated in the fiber. Finally, we obtain pure antibunched path-entangled two-photon state, and investigates its frequency entanglement property through spatial beating TPI.

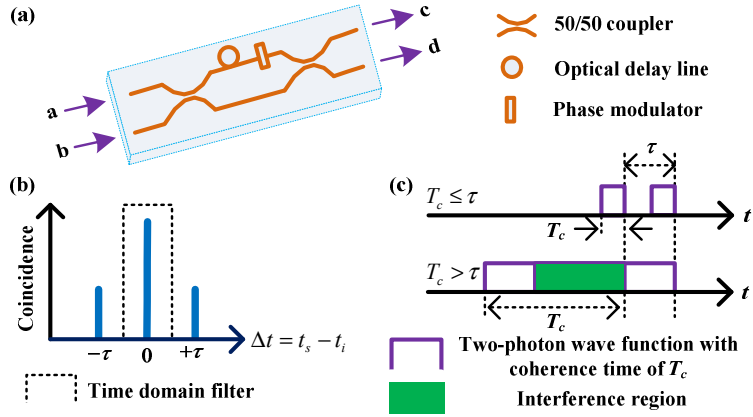

Fig. 1. The schematic of controlled entanglement router. (a) The structure of an UMZI. (b) Time-resolved two-photon coincidence measurement. (c) The interference property of selected two-photon states.

Figure 1(a) shows the structure of an UMZI, which consists of two 50/50 couplers, an optical delay line with a delay time of $\tau$ and a phase modulator introducing a relative phase difference of $\varphi$ between the two arms. The UMZI is a four ports linear optical device with two input spatial modes $a$ and $b$, and two output spatial modes $c$ and $d$. Let frequency nondegenerate energy-time entangled two-photon state $|\omega_i \omega_s\rangle_a$ be input into UMZI from spatial mode $a$, where $\omega_i$ and $\omega_s$ are angular frequencies of idler and signal photons, respectively. For simplicity the behaviors of the two 50/50 couplers, optical delay line and phase modulator in the UMZI are all assumed to be frequency independent. In such case, the two-photon state at the output of the first 50/50 coupler is,

$$|\zeta_0\rangle = \frac{1}{\sqrt{2}}\left(|\psi_1'\rangle + i|\psi_2'\rangle\right)$$

$$|\psi_1'\rangle = \frac{1}{\sqrt{2}}(|\omega_i\rangle_S|\omega_s\rangle_S - |\omega_i\rangle_L|\omega_s\rangle_L), |\psi_2'\rangle = \frac{1}{\sqrt{2}}(|\omega_i\rangle_S|\omega_s\rangle_L + |\omega_i\rangle_L|\omega_s\rangle_S)$$

(1)

where $|\psi_1'\rangle$ and $|\psi_2'\rangle$ are the bunched, i.e. two photons output from the same spatial mode simultaneously, and antibunched, i.e. two photons output from different spatial modes simultaneously, path-entangled two-photon state, respectively. $|\omega_{i,s}\rangle_S$ and $|\omega_{i,s}\rangle_L$ are the states for idler or signal photons taking the spatial modes corresponding to the long and short arms in the UMZI, respectively.

The path-entangled two-photon states output from the first 50/50 coupler naturally evolve along the two arms and meet at the second 50/50 coupler. For the state $|\psi_1'\rangle$, the two photons take the same arm together and output from the UMZI simultaneously; while for the state $|\psi_2'\rangle$, the two photons take the two arms separately and output from the UMZI with time delays of $\Delta t = t_s - t_i = \pm\tau$, where $t_s$ and $t_i$ are the output moments of signal and idler photons, respectively. The contributions of $|\psi_1'\rangle$ and $|\psi_2'\rangle$ in the output two-photon states are distinguished by time-resolved two-photon coincidence measuring. If a time domain filter is applied to select two-photon states with $\Delta t = 0$ and reject the ones with $\Delta t = \pm\tau$, the contribution of $|\psi_1'\rangle$ can be selected as shown in Fig. 1(b). It is found that half of the entanglement within the input energy-time entangled two-photon states is selected, while the other half is lost [13]. It is worth to note that the generated signal and idler photons have wide spectral widths [14, 15], hence the single photon coherence time of signal or idler photons (denoted by $\tau_{coh}$) is much smaller than $\tau$ in the UMZI, which prevents the single photon interference of idler or signal photons. The output two-photon states selected by the time domain filter result from the interference of $|\psi_1'\rangle$ at the second 50/50 coupler, due to the indistinguishability of two arms in the UMZI. The selected two-photon states can be expressed as,

$$|\zeta_1\rangle = \frac{1}{\sqrt{2}}\left(\left(1+e^{i((\omega_s+\omega_i)\tau+2\varphi)}\right)|\psi_1\rangle + i\left(1-e^{i((\omega_s+\omega_i)\tau+2\varphi)}\right)|\psi_2\rangle\right)$$

$$|\psi_1\rangle = \frac{1}{\sqrt{2}}\left(|\omega_i\rangle_c|\omega_s\rangle_c - |\omega_i\rangle_d|\omega_s\rangle_d\right), |\psi_2\rangle = \frac{1}{\sqrt{2}}\left(|\omega_i\rangle_c|\omega_s\rangle_d + |\omega_i\rangle_d|\omega_s\rangle_c\right) \quad (2)$$

where $|\omega_{i,s}\rangle_c$ and $|\omega_{i,s}\rangle_d$ represent the states with idler or signal photons output from modes $c$ and $d$, respectively; $|\psi_1\rangle$ and $|\psi_2\rangle$ are the bunched and antibunched path-entangled two-photon states output from the UMZI, respectively; $(\omega_s+\omega_i)\tau$ is a constant determined by the angular frequency $\omega_p$ of pump light. The derivation of Eq. (2) requires that the coherent time $T_c$ of the two-photon wave function of $|\psi_1'\rangle$ is far larger than $\tau$, i.e. $T_c \gg \tau$. However, $T_c$ is limited in experiment, which influences the interference of the two-photon state as illustrated in Fig. 1(c). When $T_c \leq \tau$, the two-photon wave function of $|\psi_1'\rangle$ do not interfere with itself since the wave functions through the two arms of the UMZI are distinguishable in time; while $T_c > \tau$, two-photon wave function of $|\psi_1'\rangle$ partially interfere with itself because the two-photon wave functions through the UMZI are indistinguishable in the time region from $\tau$ to $T_c$. In Eq. (2), it can be seen that the two-photon state output from the UMZI is a superposition state of bunched and antibunched path-entangled two-photon states, indicating that the selected entanglement within the input two-photon state can be coherently routed between the two path-entangled two-photon states. Similar results can be obtained while other two-photon states are input into the UMZI such as $|\omega_i\omega_s\rangle_b$, $|\omega_i\rangle_b|\omega_s\rangle_a$ and $|\omega_i\rangle_a|\omega_s\rangle_b$. The output possibilities of bunched and antibunched path-entangled two-photon states are controlled by $\varphi$, which are

expressed as,

$$P_1 = 0.5 \times (1 + \cos((\omega_s + \omega_i)\tau + 2\varphi))$$
$$P_2 = 0.5 \times (1 - \cos((\omega_s + \omega_i)\tau + 2\varphi))$$
(3)

where $P_1$ and $P_2$ are the output possibilities of $|\psi_1\rangle$ and $|\psi_2\rangle$, respectively. It can be seen that $P_1$ and $P_2$ periodically vary with $\varphi$, while the sum of them equals to 1 under any $\varphi$. Therefore the UMZI acts as a controlled entanglement router with two virtual output ports, i.e. the bunched and antibunched virtual ports, through which the selected entanglement is coherently routed between the two virtual output ports, while the routing process is controlled by $\varphi$. Utilizing such controlled entanglement router, pure bunched or antibunched path-entangled two-photon states can be directly obtained.

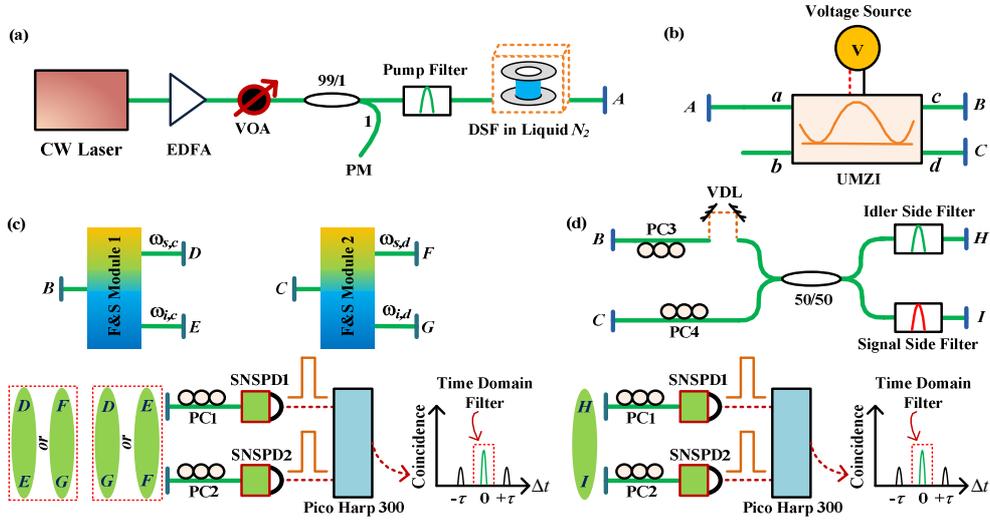

Fig. 2. Experiment setups for the demonstration of controlled entanglement router. (a) Setups for frequency nondegenerate energy-time entanglement generation; EDFA: Erbium Doped Fiber Amplifier; VOA: Variable Optical Attenuator; PM: Power Meter; DSF: Dispersion Shifted Fiber; 99/1: 99/1 coupler. (b) The structure of UMZI used in the experiment. (c) Setups for measuring nonclassical TPI; F&S: Filtering and Splitting System; SNSPD: Superconducting Nanowire Single Photon Detector; PC: Polarization Controller. (d) Setups for measuring spatial beating TPI; VDL: Variable Delay Line; 50/50: 50/50 coupler.

Figure 2(a) is the setups for generating frequency nondegenerate energy-time entangled two-photon state in fiber. A piece of dispersion shifted fiber (DSF, 800 meters in length) is pumped by a continuous wave (CW) monochromatic laser (Angilent 81980A) with a linewidth of 100 kHz. The wavelength of pump light is set as 1552.16 nm, closing to the zero dispersion wavelength $\lambda_0$=1549 nm of the DSF. To reduce the Raman noise photons generated from spontaneous Raman scattering (SpRS), the DSF is cooled in liquid nitrogen [16]. The power of pump light is amplified by an erbium doped fiber amplifier (EDFA). A variable optical attenuator (VOA) and a 99/1 fiber coupler with a power meter (PM) is used to control and monitor pump power level, which is 10 mW in the experiment. The side-band noise photons in pump light are rejected by a pump filter with a side-band rejection of greater than 120 dB. Energy-time entangled two-photon state is generated by spontaneous four-wave mixing (SFWM) process in DSF and the generated photon pair is wavelength nondegenerate. The photon with longer wavelength is called signal photon, while the one with shorter is idler photon. The emission simultaneity of photon pair is ensured by energy conservation, and the coherence property of which is guaranteed by momentum conservation in SFWM process. The coherence time of the generated two-photon state $T_c$ is about 10 μs (equaling to the coherence time of pump light estimated by its linewidth), implying that the emission moment of a photon pair is unpredictable in a 10 μs time domain window.

The generated energy-time entangled two-photon state is directly input into the UMZI from spatial mode *a*, as shown in Fig. 2(b). The UMZI used in the experiment is a part of a commercial 10 GHz differential quaternary phase-shift keying (DQPSK) demodulator for coherent optical fiber communication (Optoplex Corp., DI-CAKFASO15-R1). The total insertion loss of the device is 4.2 dB, including an addition loss of 3 dB introduced by a 50/50 splitter in it. The time delay of $\tau$ between the two arms is 100 ps. At the output end of the UMZI, the two-photon state exits from ports *B* and *C*, corresponding to the spatial modes *c* and *d* of the UMZI, respectively. Generated two-photon states through the SFWM processes in fiber have wide spectral distribution, hence the signal and idler photons satisfying the energy conservation are selected by two filtering and splitting modules (F&S Module 1 and 2) as shown in Fig. 2(c), with an insertion loss of 0.8 dB. The selected wavelengths of signal and idler photons are centered at 1555.75 and 1549.32 nm respectively, with a -3 dB spectral width of 32 GHz, respectively. A pump light isolation of greater than 120 dB is achieved in the two F&S modules. The coherent time $\tau_{coh}$ of the filtered signal and idler photons is about 30 ps estimated by the spectral width. Hence, single photon interference effect is eliminated in the UMZI due to $\tau > \tau_{coh}$. On the other hand, $\tau$ is much smaller than the $T_c$ of 10 μs, which ensures that the entanglement within generated two-photon state can be routed between the two virtual output ports of the bunched and antibunched path-entangled two-photon states. The photon states output from ports *D*, *E*, *F* and *G* correspond to $|\omega_s\rangle_c$, $|\omega_i\rangle_c$, $|\omega_s\rangle_d$ and $|\omega_i\rangle_d$ states in Eq. (2), respectively. It worth to note that the frequency nondegenerate property of the photon pair is good for path-entangled states discriminating due to the photon can be separated by filters. The entanglement routing process is controlled by $\varphi$, which is adjusted by a voltage source as shown in Fig. 2(b).

The possibilities of the entanglement outputs from the bunched and antibunched virtual ports, can be presented by the coincidence counts, which are measured by time-resolved two-photon coincidence measurement. To measure the possibility of the entanglement output from the bunched virtual port, ports *D* and *E* (or *F* and *G*) are connected to two superconducting NbN-nanowire single-photon detectors, SNSPD1 and 2, respectively; while for antibunched virtual port, ports *D* and *G* (or *E* and *F*) are connected to SNSPD1 and 2, respectively. Two polarization controllers (PC1 and 2) are placed before SNSPDs to adjust the polarization state of photons due to the detection efficiencies of the SNSPDs are polarization dependent. The SNSPD1 and 2 are installed in one Gifford-McMahon cryocooler at a working temperature of 2.2 K, with a detection efficiency of 6.0% and 4.0% under a dark count rate of 10 Hz, respectively; the timing jitters of SNSPD1 and 2 are 25 and 44 ps, respectively [17]. The output of SNSPDs are sent into a time-correlated single photon counting module (TCSPC, PicoHarp 300, PicoQuant) for time-resolved two-photon coincidence measurement, in which the used time-bin width is 4 ps. Owing to the small timing jitters of SNSPDs and the high time resolution of the TCSPC module, a high resolution time-resolved two-photon coincidence measurement can be obtained, and a very narrow time domain filter is applied to select two-photon states in the experiment.

Figure 3 shows a typical result of the time-resolved two-photon coincidence measurement. The entanglement output from the antibunched virtual port is measured with ports $E$ and $F$ connected to SNSPDs. The bars and squares are experiment results under $\varphi$ of $\pi/2$ and $\pi$, while solid and dashed lines are the Gaussian fitting envelops of them for eye guiding. Three distinct coincidence peaks are obtained with a 100 ps time delay between adjacent peaks, which corresponds to the $\tau$ in the UMZI. Every single fitting peak has a full width at half maximum (HFWM) of 35 ps, mainly caused by the timing jitters of SNSPDs. The central peak corresponds to the entanglement output from the antibunched virtual port, which is obtained by applying a time domain filter of 88 ps with a high fidelity, as shown in Fig. 3. The accidental coincidence counts caused by noise counts are measured by shifting the time domain filter out of the three coincidence peaks and calculating the sum of the background coincidence in it. It is can be calculated that the coincidence to accidental coincidence ratio (CAR) is 32±5 in the experiment. The central coincidence peak vanishes when $\varphi=\pi$, which indicates that there is no entanglement output from the antibunched virtual port, and the entanglement within the input two-photon state is coherently routed to the bunched virtual port under this condition.

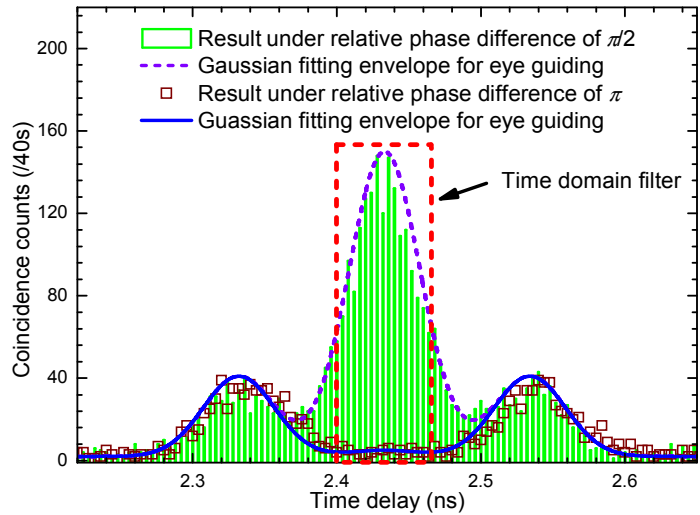

Fig. 3. Typical time-resolved two-photon coincidence measurement results. The bars and squares are experiment results under $\varphi$ of $\pi/2$ and $\pi$, respectively; dashed and solid lines are Gaussian fitting envelops for eye guiding.

To examine the coherently entanglement routing scheme, entanglement output form the bunched and antibunched virtual ports is measured under different $\varphi$ as shown in Fig. 4, in which coincidence counts are the sum over the central coincidence peak in the time domain filter of 88 ps. Figure 4(a) is the measured results output from the antibunched virtual port, measured by connecting ports $E$ and $F$ to SNSPD1 and 2, respectively. Circles are experiment results, while solid line is the fitting curve according to Eq. (3), showing a fitting visibility of $(96.9\pm1.0)\%$, without subtracting the accidental coincidence counts. Figure 4(b) is the results output from the bunched virtual port, measured by connecting ports $D$ and $E$ to SNSPD1 and 2, respectively. Squares are experiment results, while solid line is the fitting curve, also showing a fitting visibility of $(96.9\pm1.0)\%$, without subtracting the accidental coincidence counts. It can be seen that the coincidence counts in the two virtual ports vary with $\varphi$ under a period of $\pi$, while the single photon counts of signal and idler sides do not vary with $\varphi$ as shown in the insets of Fig. 4(a) and (b). The results indicate that for every virtual port, the entanglement output from it is controlled by $\varphi$; while for the two virtual ports, the routed entanglement output from the router is almost unchanged with $\varphi$, showing that the controlled entanglement router must coherently routing between the bunched and antibunched virtual ports. The visibility of the fringes indicates the off-ratio of the router achieves $14.9\pm1.8$ dB.

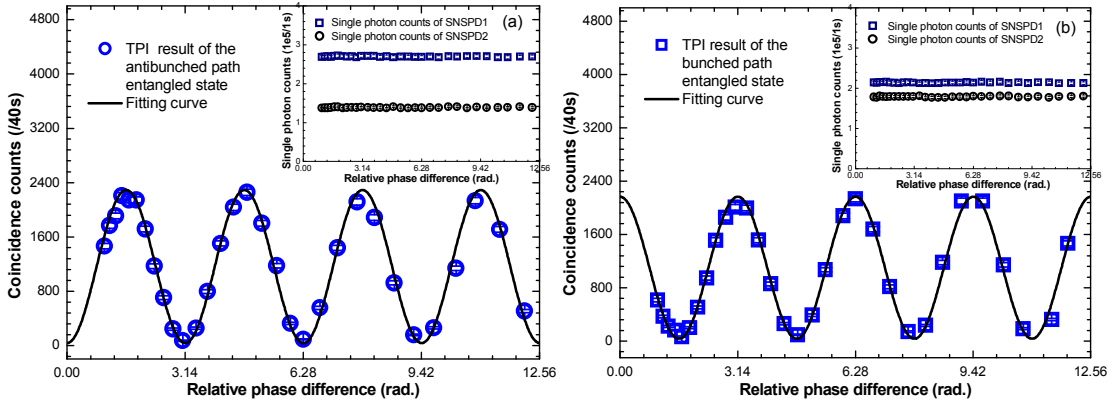

Fig. 4. Measured entanglement output from the bunched and antibunched virtual ports of the controlled entanglement router. (a) is the result of antibunched virtual port; circles are experiment results; solid line is fitting curve with a visibility of $96.9\pm1.0\%$; (b) is the result of bunched virtual port; squares are experiment results; solid line is fitting curve with a visibility of $96.9\pm1.0\%$; inserts of (a) and (b) are single side count rates.

Pure antibunched or bunched path-entangled two-photon states must output from the antibunched or bunched virtual ports of the controlled entanglement router, by setting $\varphi$ as $k\pi+\pi/2$ or $k\pi$ ($k$ is an integer), respectively. In the experiment, pure antibunched path-entangled two-photon state is obtained under $\varphi=3\pi/2$, which is the frequency-entangled two-photon state for frequency nondegenerate two-photon state input into the router [18, 19]. The frequency entanglement property of antibunched path-entangled two-photon state can be measured by the spatial beating TPI. The experiment setups for measuring the spatial beating effect are shown in Fig. 2(d). The two-photon state from ports $B$ and $C$ is input into a 50/50 fiber coupler, with the relative arrival time delay of $\Delta\tau$ between two photons is controlled by a variable delay line (VDL, MDL-002, General Photonics Corp.). PC3 and 4 are used to ensure that the input photons of 50/50 coupler are in identical polarization state. Photons from 50/50 coupler pass through signal and idler side filters and are detected by SNSPD1 and 2, respectively. The outputs of two SNSPDs are sent into time-resolved two-photon coincidence measurement system to obtain coincidence counts. The normalized coincidence counts can be expressed as [18-20],

$$P_{co} \propto 1 - V_0 \xi(\sigma \times \Delta\tau)\cos(|v_i - v_s| \times \Delta\tau)$$
$$\xi(\sigma \times \Delta\tau) = \mathrm{sinc}(\sigma \times \Delta\tau)$$
(4)

where $V_0$ is the visibility of spatial beating TPI; $\xi(\sigma \times \Delta\tau)$ is a function related to spectral properties of signal and idler photons, which is $\mathrm{sinc}(\sigma \times \Delta\tau)$ due to signal and idler side filters are super-Gaussian shaped, with an angular frequency bandwidth of $\sigma = 2\pi \times 32 \times 10^9$ in the experiment; $v_{i,s}$ are the frequency of idler and signal photons; the frequency spacing of them is 800 GHz in the experiment. Figure 5 shows results of the spatial beating TPI. Here, the coincidence counts are also obtained by summing the central coincidence peak in a time domain filter of 88 ps. Squares are experiment results, without subtracting the accidental coincidence counts. Solid line is sinusoidal fitting curve with a fitting visibility of $(99.0\pm0.8)\%$ following Eq. (4). The fidelity of the obtained antibunched path entangled two-photon state is $F=(99.5\pm0.4)\%$, calculated by $F=(1+V_0)/2$ [18].

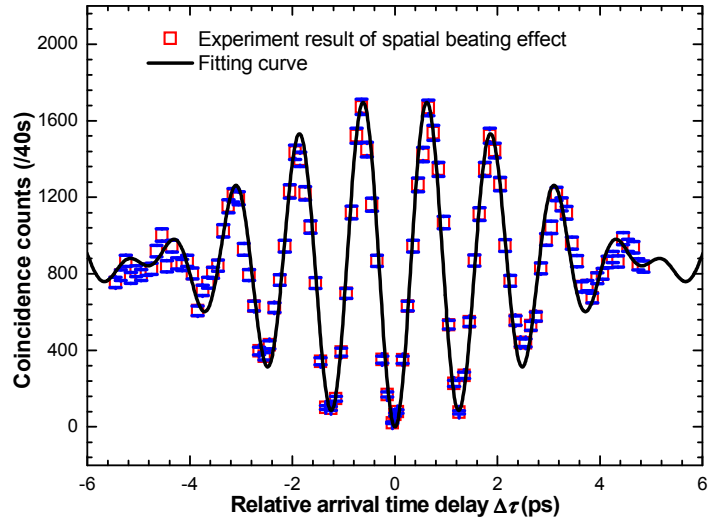

Fig. 5. Spatial beating TPI of antibunched path-entangled two-photon states; Squares are experiment results; solid line is fitting curve with a fitting visibility of $(99.0\pm0.8)\%$.

Figure 5 shows the period of spatial beating TPI fringe is 1.25 ps, i.e. a period of 375 μm in length, which is exactly determined by the frequency spacing between the idler and signal photons. Spatial beating effect also exist while bunched path-entangled two-photon state inputs into 50/50 fiber coupler, in which the normalized coincidence count is in proportion to $1 - V_0 \mathrm{sinc}(\sigma \times \Delta\tau)\cos(|v_i + v_s| \times \Delta\tau)$, resulting in a beating period of 2.59 fs, i.e. a period of 776.08 nm in length, determined by the sum of frequency of signal and idler photons. It worth to note that the spatial beating effect is a four-order interference with two-photon state, which is immune to environment disturbing due to the indistinguishable property of the path entanglement [20]. Hence, the antibunched and bunched path-entangled two-photon states are useful for very high resolution relative length measurement in disturbing environment. On the other hand, the obtained antibunched path-entangled two-photon state is entangled in freedoms of frequency, energy-time and optical path, so the controlled entanglement router is very useful for hyper-entanglement preparation, which is desired in quantum information application [21].

In summary, we have proposed and experimentally demonstrated that the UMZI can be employed as a controlled entanglement router with two virtual output ports of bunched and antibunched path-entangled two-photon states, and the routing process in the router is controlled by the relative phase difference in it. A commercial devices based experiment has been developed, in which frequency nondegenerate energy-time entanglement is generated based on SFWM process in fiber and a commercial optical communication device is used as an UMZI. Our experiment results show that half of the entanglement within the input two-photon state is periodically routed between the two virtual ports, by changing the relative phase difference in the UMZI with an off-ratio of $14.9\pm1.8$ dB. Pure antibunched path-entangled two-photon state has been obtained and verified to be a frequency-entangled two-photon state by a spatial beating TPI experiment with a visibility of $(99.0\pm0.8)\%$. In the experiment, we can generate energy-time entanglement in a silicon waveguide chip [22], while the UMZI can also be fabricated on the same silicon chip. So the generation and routing of energy-time entanglement can be realized in an integrated fashion, which is desired in quantum photonics [23].

This work is supported by 973 Programs of China under Contract Nos. 2011CBA00303 and 2010CB327606, China Postdoctoral Science Foundation, Tsinghua University Initiative Scientific Research Program, Basic Research Foundation of Tsinghua National Laboratory for Information Science and Technology (TNList), National Natural Science Foundation of China under Contract No. 91121022, Strategic Priority Research Program (B) of the Chinese Academy of Sciences under Contract Nos. XDB04010200 and XDB04020100.